\documentclass[conference]{IEEEtran}
\usepackage[latin9]{inputenc}
\usepackage{color}
\usepackage{amsmath}
\usepackage{amssymb}
\usepackage{graphicx}

\makeatletter
\IEEEoverridecommandlockouts
\usepackage{cite}
\usepackage{amsfonts}
\usepackage{algorithmic}
\usepackage{textcomp}
\usepackage{xcolor}
\def\BibTeX{{\rm B\kern-.05em{\sc i\kern-.025em b}\kern-.08em
    T\kern-.1667em\lower.7ex\hbox{E}\kern-.125emX}}

\makeatother

\begin{document}
\title{Reconfigurable Intelligent Surfaces based Cognitive Radio Networks}
\author{\textcolor{black}{\normalsize{}Abubakar U. Makarfi}\textit{\textcolor{black}{\normalsize{}$^{1}$,
}}\textcolor{black}{\normalsize{}Rupak Kharel}\textit{\textcolor{black}{\normalsize{}$^{1}$,}}\textcolor{black}{\normalsize{}
Khaled M. Rabie}\textit{\textcolor{black}{\normalsize{}$^{1}$}}\textcolor{black}{\normalsize{},
Omprakash Kaiwartya}\textit{\textcolor{black}{\normalsize{}$^{2}$}}\textcolor{black}{\normalsize{},
Xingwang Li}\textit{\textcolor{black}{\normalsize{}$^{3}$}}\textcolor{black}{\normalsize{},
Dinh-Thuan Do}\textit{\textcolor{black}{\normalsize{}$^{4}$}}\textcolor{black}{\normalsize{}}\\
\textcolor{black}{\normalsize{}$^{1}$Faculty of Science and Engineering,
Manchester Metropolitan University, UK}\\
\textit{\textcolor{black}{\normalsize{}$^{2}$}}\textcolor{black}{\normalsize{}School
of Science and Technology, Nottingham Trent University, UK }\\
\textit{\textcolor{black}{\normalsize{}$^{3}$}}\textcolor{black}{\normalsize{}School
of Physics and Electronic Information Engineering, Henan Polytechnic
University, China }\\
\textit{\textcolor{black}{\normalsize{}$^{4}$}}\textcolor{black}{\normalsize{}Faculty
of Electrical \& Electronics Engineering, Ton Duc Thang University,
Vietnam.}\\
\textcolor{black}{\normalsize{}Email:\{a.makarfi, r.kharel, k.rabie\}@mmu.ac.uk;
omprakash.kaiwartya@ntu.ac.uk; }\\
\textcolor{black}{\normalsize{}lixingwang@hpu.edu.cn; dodinhthuan@tdtu.edu.vn.}}
\maketitle
\begin{abstract}
Over the last decade, cognitive radios (CRs) have emerged as a technology
for improving spectrum efficiency through dynamic spectrum access
techniques. More recently, as research interest is shifting beyond
5G communications, new technologies such as reconfigurable intelligent
surfaces (RISs) have emerged as enablers of smart radio environments,
to further improve signal coverage and spectrum management capabilities.
Based on the promise of CRs and RISs, this paper seeks to investigate
the concept of adopting both concepts within a network as a means
of maximizing the potential benefits available. The paper considers
two separate models of RIS-based networks and analyzes several performance
metrics associated with the CR secondary user. Monte Carlo simulations
are presented to validate the derived expressions. The results indicate
the effects of key parameters of the system and the clear improvement
of the CR network, in the presence of a RIS-enhanced primary network.
\end{abstract}


\section{Introduction}

Recent research interest has shifted towards beyond 5G communications,
with new technologies such as reconfigurable intelligent surfaces
(RISs) emerging as enablers of smart radio environments, to further
improve signal coverage and spectrum management capabilities.

\textcolor{black}{The RIS concept is envisaged as a technology integral
to the realization of next generation beyond 5G and 6G networks. The
aim is to enable a controllable smart radio environment, with} improved
coverage, spectrum efficiency and signal quality \textcolor{black}{\cite{Renzo2019,Liaskos_metasurface}.
RISs are man-made surfaces of electromagnetic materials that are electronically
controlled and have unique wireless communication} capabilities \cite{Basar2019WirelessCT}.
RIS-based transmission schemes have several key features, such as
full-band response, nearly passive with dedicated energy sources,
easy deployability on different surfaces like buildings, vehicles
or indoor spaces, as well as being nearly unaffected by receiver noise
\cite{Basar2019WirelessCT,Basar_xmsn_LIS}. As a result of these benefits,
RIS-based technologies have been studied in the literature for enhancing
wireless security \cite{gong2019RISsurvey,PLSv2vRISMakarfi}, suppressing
interference \cite{Chen_2016,HollowayReview}, improving energy and
spectral efficiency \cite{EE_MU_MISO_LIS,Huang2018ReconfigurableIS,subrt}
and enhancing multi-user networks \cite{RIS_MU_MIMO,EE_MU_MISO_LIS,LIS_multi_user}.

A relatively earlier and more investigated technology for improving
spectrum efficiency is the cognitive radio (CR), employing the concept
of dynamic spectrum access \cite{zhangDSA_CRN}. A CR network is comprised
of primary users (PUs) and secondary users (SUs) co-existing within
a non-cooperating shared network. The PU is traditionally prioritized
because it is the licensed network user, while the unlicensed SU\textcolor{black}{{}
shares the spectrum on condition of non-harmful interference to the
PU. }One of the most important tasks of the secondary CR node therefore,
is spectrum sensing. Sensing involves detecting transmission opportunities
or spectrum holes, which stand for those sub-bands of the radio spectrum
that are underutilized (in part or in full) at a particular time and/or
spatial region. Notwithstanding the stated benefits of RIS-enabled
networks, in interference supression and signal-to-noise-ratio (SNR)
maximization, to the best of our knowledge, very little research has
been reported on RIS-enabled CR networks.

In this regard, recent investigations on RIS-enabled CR networks include
\cite{XuIRS_FD_CR,GuanIRS_CR,YuanRIS_CR,zhangRIS_CR}. In \cite{XuIRS_FD_CR},
the outage probability and system sum rate metrics were determined
for a CR network, with particular interest in improving spectral efficiency
through resolving the resource allocation problem in a full-duplex
secondary underlay system. In \cite{GuanIRS_CR}, the RIS technology
was adopted to reduce the challenge of decreased secondary network
throughput, due to the presence of strong cross-link interference
with the primary user. In \cite{YuanRIS_CR}, a downlink MISO CR network
with multiple PUs in a RIS-enabled network is investigated, while
in \cite{zhangRIS_CR}, a beamforming design algorithm for a RIS-enabled
CR network with imperfect channel state information was presented,
to minimze the interference on the primary network.

From the aforementioned discussion, our motivation for this study
stems from the fact that all the cited RIS-related CR literature considered
only the configuration of the RIS as a relay, while several other
RIS configurations have been studied in the literature, such as the
RIS as a receiver or as an access point at the transmitter \cite{Basar_xmsn_LIS,gong2019RISsurvey,PLSv2vRISMakarfi,RISfisherMakarfi}.
Moreover, no specific invesigations have been reported for the crucial
task of spectrum sensing by the CR node. In light of this, our main
contributions are; Firstly, we demonstrate how the CR spectrum sensing
task can be enhanced by employing RIS-based technologies at the PU.
To achieve this, we present system models for two popular RIS-configurations
existing in the literature \cite{Basar_xmsn_LIS,gong2019RISsurvey,PLSv2vRISMakarfi,RISfisherMakarfi};
the RIS relay and the RIS access point for transmission. Secondly,
we analyze the performance of the CR network. We derive expressions
for a CR node's, false alarm and detection probabilities as well as
transmission probability and throughput. We also present expressions
for the asymptotic performance of the system throughput. All expressions
were verified using Monte Carlo simulations.

The paper is organized as follows. In Section \ref{sec:Sys-Model},
we describe both system configurations under study. Thereafter, in
Section \ref{sec:Perf-anal}, we derive expressions for efficient
computation of the various performance metrics for both RIS configurations.
In Section \ref{sec:Results}, we present the results and discussions,
followed by highlights of the main conclusions in Section \ref{sec:Conclusions}.

\section{System Model\label{sec:Sys-Model}}

Consider a shared network comprising independent PUs and secondary
CR nodes operating within a spatial region. The PU has priority of
transmission within the network, while the secondary CR nodes are
allowed to access the channel only when the PU is idle. This is achieved
through a spectrum sensing protocol where a CR node measures the cumulative
interference present before transmission. The PU transmitter (tagged
PU1) is assumed to employ two different configurations of a RIS-based
scheme, to communicate directly to another PU receiver (tagged PU2).
During a transmission cycle, all nodes in the network are assumed
to be relatively stationary to not affect the relative phases to and
from the RIS. For the systems considered, we assume an intelligent
AP with the RIS having knowledge of channel phase terms, such that
the RIS-induced phases can be adjusted to maximize the received SNR
through appropriate phase cancellations and proper alignment of signals
from the intelligent surface.

As for the secondary network, we consider an arbitrary reference CR
node (tagged CR0), which has the capability to sense the spectrum
for opportunities to transmit. As earlier mentioned, a spectrum sensing
protocol is enforced, where CR0 measures the cumulative interference
present before transmission. When the received energy is below a predefined
threshold, then the user transmits. Otherwise, the user defers. The
binary hypothesis of the received signal at CR0 can be represented
as
\begin{equation}
s_{0}=\begin{cases}
w_{0}, & H_{0}\\
\vphantom{}\\
hx_{n}e^{-j\phi_{n}}+w_{0}, & H_{1}
\end{cases}\label{eq:binary hyp s0}
\end{equation}
where the hypotheses $H_{0}$ and $H_{1}$ represent the absence or
presence of the primary user signal, respectively. The term $h$ represents
the channel amplitude, $x_{n}$ represents the signal from PU1 and
$w_{0}$ the additive white Gaussian noise (AWGN) at CR0. $\phi_{n}$
in \eqref{eq:binary hyp s0} is the reconfigurable phase induced by
the $n$th reflector of the RIS with $N$ reflector elements, which
through phase matching, the SNR of the received signals can be maximized. 

The channel and signal model for the two RIS configurations under
considerations are as follows\footnote{These two RIS configuration types have been discussed extensively
in previous studies \cite{Basar_xmsn_LIS,gong2019RISsurvey}.}:
\begin{itemize}
\item Configuration I: \textit{PU transmission with Access Point RIS}. In
this configuration, PU1 is assumed to employ a RIS-based scheme in
the form of an access point to communicate over the network. The RIS
can be connected over a wired link or optical fiber for direct transmission
from PU1 and can support transmission without RF processing. The PU
signal and channel amplitude term in \eqref{eq:binary hyp s0} is
represented as $hx_{n}=\sum_{n=1}^{N}h_{c,n}$, where $h_{c,n}=\sqrt{g_{c,n}r_{c}^{-\beta_{c}}e^{-j\psi_{n}}}$
is the channel coefficient of the PU1-to-CR link with distance $r_{c}$,
path-loss exponent $\beta_{c}$, phase component $\psi_{n}$ and $g_{c,n}$
assumed to model Rayleigh distribution.
\item Configuration II: \textit{PU transmission through RIS Relay}. In this
configuration, we consider a RIS-based scheme with the RIS employed
as a relay or reflector for PU communication in the network. The RIS
can be deployed on a large surface or building. The signals from PU1
are reflected intelligently towards another PU receiver. The PU signal
and channel amplitude term in \eqref{eq:binary hyp s0} is represented
as $hx_{n}=\sum_{n=1}^{N}h_{r}h_{c,n}e^{-j\phi_{n}}$, where $h_{r}=\sqrt{g_{r,n}r_{r}^{-\beta_{r}}e^{-j\theta_{n}}}$
is the channel coefficient of the PU1-to-RIS link with distance $r_{r}$,
path-loss exponent $\beta_{r}$, phase component $\theta_{n}$ and
$g_{r}$ following a Rayleigh fading distribution. The term $h_{c,n}=\sqrt{g_{c,n}r_{c}^{-\beta_{c}}e^{-j\psi_{n}}}$,
is the channel coefficient from RIS-to-CR0, with distance $r_{c}$,
path-loss exponent $\beta_{c}$, phase component $\psi_{n}$ and $g_{c,n}$
also assumed to model Rayleigh distribution. 
\end{itemize}
Without loss of generality, we denote the power spectral density of
the AWGN as $N_{0}$ and assume the path-loss exponents $\beta_{r}=\beta_{c}=\beta$.
It is worth noting that, as a result of the match phasing operation,
CR0 is unlikely to receive a maximized signal. However, the received
signal is maximized when the phase functions $\phi_{n}=\psi_{n}$
(for the access point RIS), or $\phi_{n}=\theta_{n}+\psi_{n}$ (for
the RIS relay) for $n=1,2,\ldots N$. In what follows, the performance
analysis is considered, based on the aforementioned RIS configurations.

\section{Performance Analysis \label{sec:Perf-anal}}

In this section, we derive analytical expressions for some performance
measures of a CR node within the network. The performance is analyzed
in terms of the false alarm probability, detection probability, error
probability and throughput of the system. 

\subsection{Probability of False Alarm Analysis}

For a CR node, a key performance metric lies in the ability to correctly
decide the presence or absence of the PU transmission. When a CR sensing
node incorrectly decides that the PU transmission is ongoing (while
it is not, under $H_{0}$), this may result in wasted transmission
opportunities for the secondary CR node. This is referred to as the
probability of false alarm ($P_{f}$) and is generally defined with
respect to the binary hypothesis \eqref{eq:binary hyp s0} as

\begin{equation}
P_{f}=\textrm{Pr}\left(y_{0}>y_{\textrm{th}}\mid H_{0}\right)\label{eq:Pfstatement}
\end{equation}
where $y_{0}$ is the energy of the received signal $s_{0}$ at CR0
and $y_{\textrm{th}}$ a pre-determined threshold adopted to avoid
interfering with the PU. 

Under $H_{0}$, the equivalent noise variance of the AWGN process
with flat power spectral density (PSD), is a sum of the square of
a zero-mean complex Gaussian random variable (RV). Therefore, under
$H_{0}$, $y_{0}$ follows a Chi-square distribution with one degree
of freedom and PDF given by
\begin{equation}
f\left(y_{0}\right)=\frac{1}{y_{0}\Gamma\left(1/2\right)}\left(\frac{y_{0}}{N_{0}}\right)^{\frac{1}{2}}e^{-\frac{y}{N_{0}}}\label{eq:pdf y}
\end{equation}
where $\Gamma\left(.\right)$ represents the Gamma function \cite[Eq. (8.310)]{book2}
and $N_{0}$ is the noise variance under $H_{0}$. Therefore, from
\eqref{eq:Pfstatement} and \eqref{eq:pdf y}, the false alarm probability
can be obtained as 
\begin{align}
P_{f} & =\frac{\Gamma\left(\frac{1}{2},\frac{y_{\textrm{th}}}{2N_{0}}\right)}{\Gamma\left(\frac{1}{2}\right)}\nonumber \\
 & \stackrel{\left(a\right)}{=}\textrm{erfc\ensuremath{\left(\sqrt{\frac{y_{\textrm{th}}}{2N_{0}}}\right)}},\label{eq:pf-final}
\end{align}
where $\Gamma\left(a,b\right)$ represents the incomplete Gamma function
\cite[Eq. (6.5.3)]{abramowitz_stegun}, $\textrm{erfc}\left(x\right)=1-\textrm{erf}\left(x\right)$
is the complementary error function \cite[Eq. (8.250.4)]{book2} and
$\left(a\right)$ in \eqref{eq:pf-final} was obtained using the fact
that $\Gamma\left(\frac{1}{2}\right)=\sqrt{\pi}$ \cite[Eq. (8.338.2)]{book2}
and $\Gamma\left(\frac{1}{2},b\right)=\sqrt{\pi}\textrm{erfc\ensuremath{\left(\sqrt{b}\right)}}$
\cite[Eq. (6.5.17)]{abramowitz_stegun}.

\subsection{Probability of Detection Analysis}

In this section, we derive expressions for the probability of detection
for the two RIS configurations under consideration. The probability
of detection is the probability that a CR sensing node correctly determines
that the PU transmission is active. This performance indicator is
important in order to ensure that a CR node avoids causing harmful
interference to the legacy PU node. The probability of detection ($P_{d}$)
is defined with respect to the binary hypothesis \eqref{eq:binary hyp s0}
as 

\begin{equation}
P_{d}=\textrm{Pr}\left(\gamma_{0}>y_{\textrm{th}}\mid H_{1}\right),\label{eq:Pd statement}
\end{equation}
where $y_{0}$ is the energy of the received signal $s_{0}$ at CR0
and $y_{\textrm{th}}$ a pre-determined threshold adopted to avoid
interfering with the PU. Next we consider the detection probability
under the different scenarios.

\subsubsection{Detection Probability with Access Point RIS\label{subsec:Pd-ap-ris}}

When the PU node employs a RIS as an access point for transmission,
the received SNR at CR0 is rarely maximized due to the match phasing
operation. With this in mind, the instantaneous SNR at CR0 is given
by
\begin{align}
\gamma_{0} & \leq\sum_{n=1}^{N}\frac{p_{s}g_{c,n}r_{c}^{-\beta}}{N_{0}}e^{j\left(\phi_{n}-\psi_{n}\right)}\nonumber \\
 & \stackrel{\left(b\right)}{=}\sum_{n=1}^{N}\bar{\gamma}g_{c,n}r_{c}^{-\beta},\label{eq:snr0-AP}
\end{align}
where $\bar{\gamma}=\frac{p_{s}}{N_{0}}$, $p_{s}$ is the PU transmit
power and $N_{0}$ is the noise power. Line $(b)$ in \eqref{eq:snr0-AP}
can be maximized for $\gamma_{0}$ when the phase functions $\phi_{n}=\psi_{n}$
for $n=1,2,\ldots N$. 

From \eqref{eq:Pd statement} and \eqref{eq:snr0-AP}, $P_{d}$ becomes

\begin{align}
P_{d}^{\textrm{ap}} & =\textrm{Pr}\left(\gamma_{0}>y_{\textrm{th}}\right)\nonumber \\
 & =\textrm{Pr}\left(\sum_{n=1}^{N}\bar{\gamma}g_{c,n}r_{c}^{-\beta}>y_{\textrm{th}}\right)\nonumber \\
 & =\textrm{Pr}\left(\sum_{n=1}^{N}g_{c,n}>\frac{y_{\textrm{th}}}{\bar{\gamma}r_{c}^{-\beta}}\right),\label{eq:pd-AP-step2}
\end{align}
where the RV $g_{c}$ follows an independent Rayleigh distribution
with PDF given by 
\begin{equation}
f\left(x\right)=x\exp\left(-\frac{x^{2}}{2}\right).\label{eq:pdf-ray}
\end{equation}

For sufficiently large $N\gg1$, using the central limit theorem,
it can be shown that $\sum_{n=1}^{N}g_{c,n}$ can be approximated
by a Gaussian distribution. Thus, from \eqref{eq:pd-AP-step2}, the
detection probability can be represented as
\begin{equation}
P_{d}^{\textrm{ap}}\approx\frac{1}{2}\left(1+\textrm{erf}\left(\frac{\varTheta_{\textrm{ap}}-\mu_{\textrm{ap}}}{\sqrt{2\sigma_{\textrm{ap}}^{2}}}\right)\right),\label{eq:pd-AP-step3}
\end{equation}
where $\textrm{erf}\left(x\right)=\frac{2}{\sqrt{\pi}}\int_{0}^{x}e^{-t^{2}}dt$
is the error function \cite[Eq. (8.250.1)]{book2} and $\varTheta_{\textrm{ap}}=\frac{y_{\textrm{th}}}{\bar{\gamma}r_{c}^{-\beta}}$.

Using \eqref{eq:pdf-ray}, the mean ($\mu)$ and variance ($\sigma^{2}$)
of the equivalent Gaussian distribution parameters are given by
\begin{align}
\mu_{\textrm{ap}} & =\int_{0}^{\infty}\sum_{n=1}^{N}g_{c}f(g_{c})\textrm{d}g_{c}\nonumber \\
 & =\sum_{n=1}^{N}\int_{0}^{\infty}g_{c}^{2}\exp\left(-\frac{g_{c}^{2}}{2}\right)\textrm{d}g_{c}\nonumber \\
 & =N\left(\frac{\pi}{2}\right)^{\frac{1}{2}},\label{eq:mean-CLT-ap}
\end{align}
where the last step in \eqref{eq:mean-CLT-ap} was obtained using
\cite[Eq. (3.326.2)]{book2}. The variance $\sigma_{\textrm{ap}}^{2}$
is given by
\begin{align}
\sigma_{\textrm{ap}}^{2} & =\textrm{Var}\left[\sum_{n=1}^{N}g_{c,n}\right]\nonumber \\
 & =\mathbb{E}\left[\sum_{n=1}^{N}g_{c,n}^{2}\right]-\mathbb{E}\left[\sum_{n=1}^{N}g_{c,n}\right]^{2}\nonumber \\
 & =N\left(2-\frac{\pi}{2}\right).\label{eq:var-CLT-ap}
\end{align}

Therefore, from \eqref{eq:pd-AP-step3}, \eqref{eq:mean-CLT-ap} and
\eqref{eq:var-CLT-ap}, the detection probability is given by
\begin{equation}
P_{d}^{\textrm{ap}}\approx\frac{1}{2}\left(1+\textrm{erf}\left(\frac{\frac{y_{\textrm{th}}}{\bar{\gamma}r_{c}^{-\beta}}-N\left(\frac{\pi}{2}\right)^{\frac{1}{2}}}{\sqrt{2N\left(2-\frac{\pi}{2}\right)}}\right)\right).\label{eq:pd-AP-step4}
\end{equation}

\subsubsection{Detection Probability Through RIS Relay}

In this section, we derive expressions for the probability of detection
when the CR node receives a PU transmission via a reflected RIS signal.
The instantaneous SNR at CR0 is given by
\begin{align}
\gamma_{0} & \leq\sum_{n=1}^{N}\frac{p_{s}g_{r,n}g_{c,n}r_{c}^{-\beta}r_{r}^{-\beta}}{N_{0}}e^{j\left(\phi_{n}-\theta_{n}-\psi_{n}\right)}\nonumber \\
 & \stackrel{\left(d\right)}{=}\sum_{n=1}^{N}\bar{\gamma}g_{r,n}g_{c,n}\left(r_{c}r_{r}\right)^{-\beta},\label{eq:snr0-relay}
\end{align}
where $\bar{\gamma}=\frac{p_{s}}{N_{0}}$, $p_{s}$ is the PU transmit
power and $N_{0}$ is the noise power. Line $(d)$ in \eqref{eq:snr0-relay}
can be maximized for $\gamma_{0}$ when the phase functions $\phi_{n}=\theta_{n}+\psi_{n}$
for $n=1,2,\ldots N$. 

Using similar to analysis to the derivation in \ref{subsec:Pd-ap-ris},
from \eqref{eq:Pd statement} and \eqref{eq:snr0-relay}, $P_{d}$
becomes

\begin{align}
P_{d}^{\textrm{rel}} & =\textrm{Pr}\left(\gamma_{0}>y_{\textrm{th}}\right)\nonumber \\
 & =\textrm{Pr}\left(\sum_{n=1}^{N}\bar{\gamma}g_{r,n}g_{c,n}\left(r_{c}r_{r}\right)^{-\beta}>y_{\textrm{th}}\right)\nonumber \\
 & =\textrm{Pr}\left(\sum_{n=1}^{N}g_{r,n}g_{c,n}>\frac{y_{\textrm{th}}}{\bar{\gamma}\left(r_{c}r_{r}\right)^{-\beta}}\right),\label{eq:pd-relay-step2}
\end{align}
where the RVs $g_{r}$ and $g_{c}$ are independently Rayleigh distributed.
By employing the central limit theorem for the sum of RVs $\sum_{n=1}^{N}g_{r,n}g_{c,n}$,
for large $N$, then using similar analysis to the derivation in \ref{subsec:Pd-ap-ris},
the detection probability is given by
\begin{equation}
P_{d}^{\textrm{rel}}\approx\frac{1}{2}\left(1+\textrm{erf}\left(\frac{\varTheta_{\textrm{rel}}-\mu_{\textrm{rel}}}{\sqrt{2\sigma_{\textrm{rel}}^{2}}}\right)\right),\label{eq:pd-relay-step3}
\end{equation}
where $\varTheta_{\textrm{rel}}=\frac{y_{\textrm{th}}}{\bar{\gamma}\left(r_{c}r_{r}\right)^{-\beta}}$.

Using \eqref{eq:pdf-ray} and \cite[Eq. (3.326.2)]{book2}, the parameter
$\mu_{\textrm{rel}}$ in \eqref{eq:pd-relay-step3} can be obtained
as
\begin{align}
\mu_{\textrm{rel}} & =\int_{0}^{\infty}\int_{0}^{\infty}\sum_{n=1}^{N}g_{r}g_{c}f(g_{r})f(g_{c})\textrm{d}g_{r}\textrm{d}g_{c}\nonumber \\
 & =\sum_{n=1}^{N}\int_{0}^{\infty}\int_{0}^{\infty}g_{r}^{2}g_{c}^{2}\exp\left(-\frac{g_{r}^{2}+g_{c}^{2}}{2}\right)\textrm{d}g_{r}\textrm{d}g_{c}\nonumber \\
 & =N\frac{\pi}{2}.\label{eq:mean-CLT-relay}
\end{align}

From \eqref{eq:pdf-ray} and \cite[Eq. (3.326.2)]{book2}, the parameter
$\sigma_{\textrm{rel}}^{2}$ is given by

\begin{align}
\sigma_{\textrm{rel}}^{2} & =\mathbb{E}\left[\sum_{n=1}^{N}\left(g_{r}g_{c}\right)^{2}\right]-\mathbb{E}\left[\sum_{n=1}^{N}g_{r}g_{c}\right]^{2}\nonumber \\
 & =N\left(4-\frac{\pi^{2}}{4}\right).\label{eq:var-CLT-relay}
\end{align}

Thus, from \eqref{eq:pd-relay-step3}, \eqref{eq:mean-CLT-relay}
and \eqref{eq:var-CLT-relay}, the detection probability for the RIS
relay model is given by
\begin{equation}
P_{d}^{\textrm{rel}}\approx\frac{1}{2}\left(1+\textrm{erf}\left(\frac{\frac{y_{\textrm{th}}}{\bar{\gamma}\left(r_{c}r_{r}\right)^{-\beta}}-N\frac{\pi}{2}}{\sqrt{2N\left(4-\frac{\pi^{2}}{4}\right)}}\right)\right).\label{eq:pd-relay-step4}
\end{equation}

\subsection{Throughput Analysis}

In this section, we present expressions for the throughput of the
secondary network. 

The throughput can be expressed as the transmission rate per unit
area, which is a function of the node density and the detected transmission
opportunities by the CRs. If we denote the node density of the active
secondary network as $\lambda$, then the total achievable transmission
rate in the network is 
\begin{equation}
T=\lambda R_{s}P_{t},\label{eq: throughput-step1}
\end{equation}
where $P_{t}$ is the transmission probability of the reference CR
node and $R_{s}$ is the transmission rate. Ideally, $P_{t}$ can
be defined as a binary decision, such that 
\begin{equation}
P_{t}\left(\gamma_{0}\right)=\begin{cases}
1\quad & \mbox{if }\gamma_{0}\geq y_{\textrm{th}},\\
0\quad & \mbox{if }\gamma_{0}<y_{\textrm{th}},
\end{cases}\label{eq:pt-step1}
\end{equation}
where $\gamma_{0}$ is the SNR at CR0 and $y_{\textrm{th}}$ a pre-determined
threshold adopted to avoid interfering with the PU. However, in practice,
the decision to become active is modified, such that CR0 becomes active
either when the node correctly decides the PU absence under $H_{0}$
(i.e. $\Pr\left(y_{0}<y_{\textrm{th}}\mid H_{0}\right)=1-P_{f}$)
or failed to detect the presence of the PU under $H_{1}$ (i.e. $\Pr\left(y_{0}<y_{\textrm{th}}\mid H_{1}\right)=1-P_{d}$).
Thus, the transmission probability can be represented as 

\begin{equation}
P_{t}=\alpha P_{m}+\left(1-\alpha\right)\left(1-P_{f}\right),\label{eq:pt-step2}
\end{equation}
where $0<\alpha<1$ is the probability of the PU activity (or fraction
of time the PU is active) and $P_{m}=1-P_{d}$ is the probability
of a missed detection.

\begin{figure}
\begin{centering}
\includegraphics[scale=0.45]{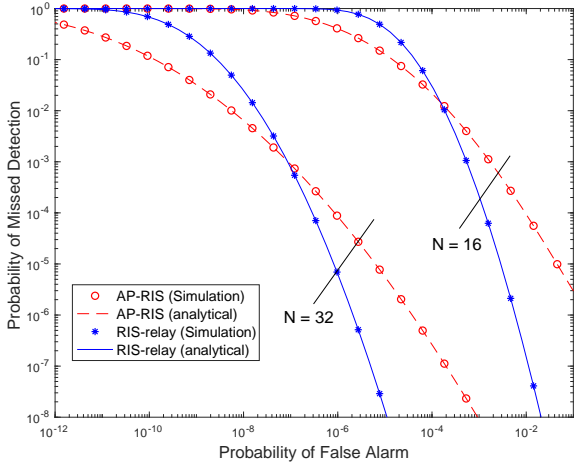}
\par\end{centering}
\caption{Complementary ROC curves for access point (AP) RIS configuration and
RIS-relay configuration. \label{fig:cr-ris-c-roc}}
\end{figure}

When the PU employs a RIS access point, then from \eqref{eq:pf-final},
\eqref{eq:pd-AP-step4}, \eqref{eq: throughput-step1} and \eqref{eq:pt-step2},
we obtain the total throughput as
\begin{align}
T_{\textrm{ap}} & =\lambda R_{s}\left\{ \alpha P_{m}^{\textrm{ap}}+\left(1-\alpha\right)\left(1-P_{f}\right)\right\} \nonumber \\
 & =\lambda R_{s}\left\{ \alpha\textrm{erf}\left(\frac{\frac{y_{\textrm{th}}}{\bar{\gamma}r_{c}^{-\beta}}-N\left(\frac{\pi}{2}\right)^{\frac{1}{2}}}{\sqrt{2N\left(2-\frac{\pi}{2}\right)}}\right)\right.\nonumber \\
 & \qquad\qquad\qquad\qquad\left.+\left(1-\alpha\right)\textrm{erf\ensuremath{\left(\sqrt{\frac{y_{\textrm{th}}}{2N_{0}}}\right)}}\right\} ,\label{eq:T-ap-final}
\end{align}
where we have used the fact that $\textrm{erf}\left(x\right)=1-\textrm{erfc}\left(x\right)$
\cite[Eq. (8.250.4)]{book2}. 

Similarly, when the CR node receives a PU transmission via a reflected
RIS signal, then from \eqref{eq:pf-final}, \eqref{eq:pd-relay-step4},
\eqref{eq: throughput-step1} and \eqref{eq:pt-step2}, we obtain
the total throughput as
\begin{align}
T_{\textrm{rel}} & =\lambda R_{s}\left\{ \alpha P_{m}^{\textrm{rel}}+\left(1-\alpha\right)\left(1-P_{f}\right)\right\} \nonumber \\
 & =\lambda R_{s}\left\{ \alpha\textrm{erf}\left(\frac{\frac{y_{\textrm{th}}}{\bar{\gamma}\left(r_{c}r_{r}\right)^{-\beta}}-N\frac{\pi}{2}}{\sqrt{2N\left(4-\frac{\pi^{2}}{4}\right)}}\right)\right.\nonumber \\
 & \qquad\qquad\qquad\qquad\left.+\left(1-\alpha\right)\textrm{erf\ensuremath{\left(\sqrt{\frac{y_{\textrm{th}}}{2N_{0}}}\right)}}\right\} ,\label{eq:T-relay-final}
\end{align}

It is worth noting that given that for the RIS configurations, where
the SNR becomes maximized, then as $N\rightarrow\infty,$ the detection
threshold $y_{\textrm{th}}$ becomes very large, such that the second
error function (in \eqref{eq:T-ap-final} and \eqref{eq:T-relay-final})
becomes $\lim{}_{\substack{x\rightarrow\infty}
}\textrm{erf}\left(x\right)=1$. It can therefore be shown that the asymptotic throughput of the
CR node for the RIS access point configuration and the RIS relay configuration
are respectively given by

\begin{align}
T_{\textrm{ap}}^{\textrm{asym}} & =\lambda R_{s}\left\{ 1-\alpha+\alpha\textrm{erf}\left(\frac{\frac{y_{\textrm{th}}}{\bar{\gamma}r_{c}^{-\beta}}-N\left(\frac{\pi}{2}\right)^{\frac{1}{2}}}{\sqrt{2N\left(2-\frac{\pi}{2}\right)}}\right)\right\} \label{eq:T-ap-asym}
\end{align}
\begin{figure}
\begin{centering}
\includegraphics[scale=0.45]{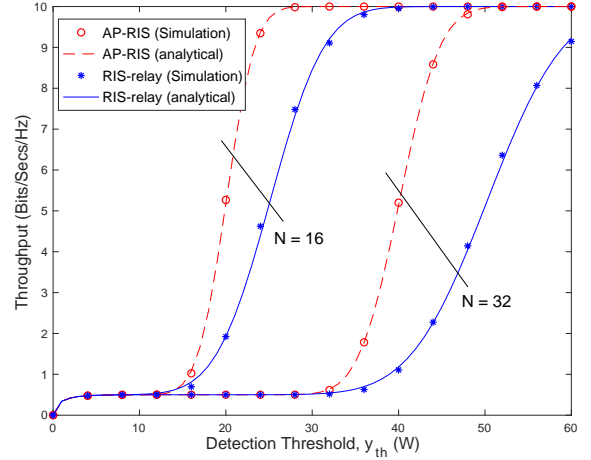}
\par\end{centering}
\caption{Throughput against detection threshold for both access point (AP)
RIS configuration and RIS-relay configuration, showing different number
of RIS cells $N$.\label{fig:throughput}}
\end{figure}
\begin{figure}
\begin{centering}
\includegraphics[scale=0.45]{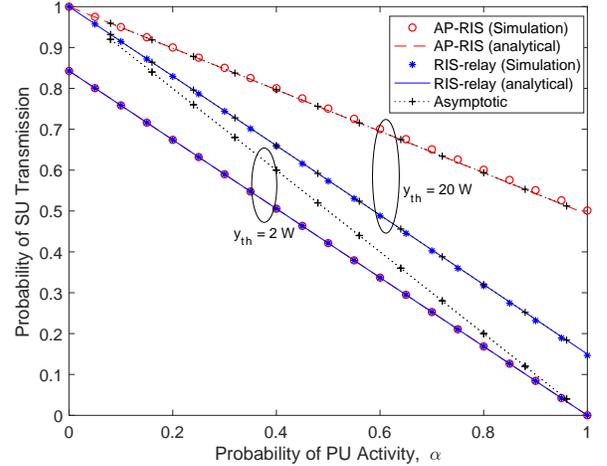}
\par\end{centering}
\caption{Probability of SU transmission against Probability of PU activity
$\alpha$ for both access point RIS configuration and RIS-relay configuration.
The asymptotic expressions are also indicated for different detection
thresholds. RIS Cells $N=16.$\label{fig:pt}}
\end{figure}
\begin{align}
T_{\textrm{rel}}^{\textrm{asym}} & =\lambda R_{s}\left\{ 1-\alpha+\alpha\textrm{erf}\left(\frac{\frac{y_{\textrm{th}}}{\bar{\gamma}\left(r_{c}r_{r}\right)^{-\beta}}-N\frac{\pi}{2}}{\sqrt{2N\left(4-\frac{\pi^{2}}{4}\right)}}\right)\right\} .\label{eq:T-relay-asym}
\end{align}

\section{Numerical Results and Discussions\label{sec:Results}}

In this section, we present and discuss some results from the mathematical
expressions derived in the paper. We then investigate the effect of
key parameters on the system. The results are then verified using
Monte Carlo simulations with at least $10^{5}$ iterations. Unless
otherwise stated, we have assumed SNR $\bar{\gamma}=0$dB, PU activeness
factor $\alpha=0.95$, $R_{s}=10$bits/secs/Hz, density $\lambda=1$
user/$m^{2}$ and all distances PU-to-RIS and RIS-to-CR normalized
to unity. One benefit of normalizing the distances is to enable a
direct comparison between the two configurations, as will be seen. 

In Fig. \ref{fig:cr-ris-c-roc}, we plot the complementary receiver
operating characteristics curve for both access point RIS and RIS-relay
configurations. The analytical curves for $P_{f}$, $P_{m}^{\textrm{ap}}$
and $P_{m}^{\textrm{rel}}$, were plotted using the expressions from
\eqref{eq:pf-final}, \eqref{eq:pd-AP-step4} and \eqref{eq:pd-relay-step4},
respectively, where $P_{m}=1-P_{d}$. We first note that the larger
number of RIS cells in both configuration results in a better performance.
In fact, it is possible to observe up to 3 to 4 orders of magnitude
performance gain, when $N$ is doubled. Furthermore, it can be observed
that the RIS-relay configuration performance better only up to about
$10^{-3}$ point for the missed-detection range, while the AP-RIS
performance better afterwards. 

In Fig. \ref{fig:throughput}, we present SU throughput results for
both access point RIS and RIS-relay configurations. The analytical
results for $T_{\textrm{ap}}$ and $T_{\textrm{rel}}$, were plotted
from \eqref{eq:T-ap-final} and \eqref{eq:T-relay-final}, respectively.
Again we observe that increasing the number of RIS cells improves
the throughput at any given detection threshold. Additionally, the
results indicate that the RIS-relay configuration outperforms the
access point RIS configuration. It is however worth noting that, an
increased detection threshold, only indicates that the CR sensing
node is willing to accept higher interference risk to the PU network,
since a higher threshold means a higher likelihood of false alarm
errors.

Next, in Fig. \ref{fig:pt}, we present a plot for the probability
of SU transmission $P_{t}$ against the probability of PU activity
$\alpha$ for both access point RIS and RIS-relay configurations ($N=16)$,
as well as different detection threshold values. The asymptotic expressions
for access point RIS and RIS-relay configurations, were plotted from
\eqref{eq:T-ap-asym} and \eqref{eq:T-relay-asym}, respectively,
by assuming $\lambda=1$ and $R_{s}=1$. Here we observe that within
the region of low detection threshold, where both RIS configurations
provide similar SU transmission probabilities, the asymptotic expressions
converge to the exact expressions, when PU activity is higher. On
the other hand, when the detection threshold is higher, it can easily
be observed that the asymptotic expressions converge to the exact
expressions, even for very low PU activities. It can therefore be
concluded that, due to the maximized SNR, only one hypothesis i.e.
$H_{1}$ is enough for the SU to make a determination to transmit,
while using both RIS configurations. It is also worth mentioning that,
for a threshold as low as $y_{\textrm{th}}=5$W, the asymptotic and
exact expressions converge. This has however been omitted in Fig.
\ref{fig:pt}, to improve clarity of the curves.

\section{Conclusions\label{sec:Conclusions}}

In this paper, we examined two popular configurations of RIS deployment
as could be applied to a CR network. We derived expressions for the
false alarm and detection probabilities as well as exact and asymptotic
expressions for the SU transmission probability and throughput of
the system. The results indicate how the use of RIS configurations
by the PU can improve the task of spectrum sensing for the secondary
CR and thereby improve the CR network performance. Further insights
were obtained by observing the asymptotic case, where it was demonstrated
that the CR could determine a transmission opportunity with a single
hypothesis, due to the maximized SNR and higher detection thresholds
allowed in the system.

\bibliographystyle{IEEEtran}
\bibliography{bibGC19}

\end{document}